\documentclass[showpacs,preprintnumbers,amsmath,amssymb,APSl,prd,nofootinbib,superscriptaddress, onecolumn]{revtex4}
\usepackage[dvips]{graphicx}
\usepackage{amssymb}
\usepackage{amsmath}
\usepackage{graphicx,color}
\usepackage{amsfonts}
\usepackage{bm}
\usepackage{cancel}
\usepackage{comment}
\usepackage{hyperref}
\usepackage{babel}

\begin{document}

\title{Regularizing Rotating Black Strings: a new black bounce solution}

\author{A. Lima}
\affiliation{Universidade Federal do Cear\'{a}, Fortaleza, Cear\'{a}, Brazil}
\email{arthur.lima@fisica.ufc.br}

\author{G. Alencar}
\email{geova@fisica.ufc.br}
\affiliation{Universidade Federal do Cear\'{a}, Fortaleza, Cear\'{a}, Brazil}

 \author{Diego~S\'aez-Chill\'on~G\'omez}
\email{diego.saez@uva.es} 
\affiliation{Department of Theoretical, Atomic and Optical
Physics and IMUVA, Campus Miguel Delibes, \\ University of Valladolid UVA, Paseo Bel\'en, 7, 47011
Valladolid, Spain}


\begin{abstract}

The present paper is devoted to a new black bounce solution that regularize the well-known rotating black string in $3+1$ dimensions. To do so, the procedure pointed out by Simpson-Visser is followed, which has been already applied successfully  to other static cases of black strings, with and without electric charge. This method implies to force a bounce on the radial coordinate, such that a wormhole throat arises before the singularity, which renders a regular solution. An analysis of the metric is conducted, showing the interpolation between a regular black hole and a wormhole, what provides a much richer family of solutions than the original metric. Different curvature magnitudes are obtained in order to analyze the regularity of the solution, including the Ricci and Kretschmann scalars. Finally, by following the Einstein field equations the corresponding effective energy-momentum tensor is obtained and the energy conditions are analyzed. 
\end{abstract}

\maketitle

\section{Introduction}\label{Intro}

Black holes arise naturally as solutions in Einstein's General Relativity (GR), featured by the presence of null hypersurfaces known as event horizons that impose a one-way path for any type of matter including light. These solutions play a crucial role in GR as they define the geometry resulting from the critical collapse of massive bodies such as stars or star clusters \cite{Joshi}. Over the last years, these solutions have gained significant attention due to technological advancements, which have led to the detection of gravitational waves by LIGO/VIRGO, as well as the first images surrounding  ``hypothetical'' supermassive black holes at the center of our galaxy and M87 galaxy \cite{LIGOScientific:2016aoc, LIGOScientific:2017vwq, EventHorizonTelescope:2019dse, EventHorizonTelescope:2022wkp}.

In GR, black holes solutions belong to the well-known Kerr-Newman family. These solutions are characterized by four basic parameters: mass $M$, angular momentum $J$, electric charge $Q$ and cosmological constant $\Lambda$ (if any). They exhibit axial symmetry (axisymmetric) and can be asymptotically flat (whether the cosmological constant becomes zero), de Sitter ($\Lambda>0$), or anti-de Sitter ($\Lambda<0$). Actually the presence of the cosmological constant influences directly their asymptotic behavior \cite{Lemos:1994xp}.  

Another type solutions own cylindrical symmetry. This type of solutions are particularly relevant in cosmology, since allow the study of topologically stable defects such as cosmic strings that may have formed during phase transitions after the \textit{big bang} \cite{Vilenkin:1984ib}. Additionally, black strings arise in the context of extra dimensions, where they represent  solutions of a p-brane model with $p=1$, where the black string corresponds to a D1-brane. In the scenario where the Universe is described by a brane in a space with extra dimensions, gravitational collapse can lead to the formation of a black hole on the brane, such that black strings are solutions to Einstein's equations in higher dimensions \cite{Muniz:2022otq}.

Black holes solutions in GR exhibit spacetime singularities, where classical concepts break down and the theory becomes non-predictable.  Singularities play a significant role when considering the evaporation of black holes through Hawking radiation, especially in the final stages of the evaporation process and it is crucial to  describe accurately the spacetime in these conditions to address fundamental questions regarding the interplay between General Relativity and Quantum Mechanics \cite{Carballo-Rubio:2018pmi}. However, it is possible to construct black holes solutions that are free of singularities \cite{Bejarano:2017fgz}. One approach is to consider spherically symmetric, static, asymptotically flat metrics with regular centers. These regular solutions provide alternative descriptions of black holes that circumvent the issues associated with singularities and contribute to a better understanding of black hole evaporation problems \cite{Hayward:2006}. In this context, several works have been published proposing regular solutions, such as the pioneering work by Bardeen \cite{bardeen:1968non}, as well as others \cite{Bambi:2013ufa, Bronnikov:2005gm, Bronnikov:2000vy}. Here, we intend to emphasize on the regular metric proposed by Simpson and Visser \cite{Simpson:2018tsi}, where the metric turns out free of singularities by introducing a modification on the Schwarzschild metric, which shows up as follows:
\begin{equation}\label{1}
    ds^2=-\left(1-\frac{2m}{\sqrt{r^2+a^2}}\right)dt^2+\frac{dr^2}{\left(1-\frac{2m}{\sqrt{r^2+a^2}}\right)}+(r^2+a^2)d\Omega^2,
\end{equation}
where "$a$" is a free parameter and "$m$" is a constant related to the mass of the central object. Here, essentially the change $r^2\rightarrow r^2+a^2$ was applied to the Schwarzschild solution. In summary, this metric exhibits the following properties: if $a>2m$, we have a two-way Morris-Thorne wormhole \cite{Simpson:2018tsi, Morris:1988cz}; for $a=2m$, a one-way wormhole arises with a horizon located at the ``throat'', similar to a Schwarzschild wormhole \cite{Simpson:2018tsi, Morris:1988cz}; and for $a<2m$, the metric describes a regular black hole with event horizons located at $r_H=\pm \sqrt{(2m)^2-a^2}$ \cite{Simpson:2018tsi}. It is worth to note that the Schwarzschild solution is recovered for $a=0$. An analysis of the tensor and curvature invariants revealed the absence of singularities at the origin for $a\neq 0$. Nevertheless, the solution requires the violation of the energy conditions, since it requires exotic sources to generate such type of solution, regardless of whether it corresponds to a black hole or a wormhole, at every point of the spacetime. In addition, the expressions for the Hawking temperature and the circular orbits for photons (photons sphere) and massive particles (ISCO) have been analyzed, where similar results in comparison to those of the Schwarzschild solution were found, despite a correcting factor of the type $\sqrt{1-a^2/k^2}$, with $k$ a constant multiple of $m$ \cite{Simpson:2018tsi}. 

Moreover, other similar solutions inspired by (\ref{1}) have been proposed in recent years. All these solutions have in common that depart from a non-regular solution and follow a similar procedure as in the Simpson-Visser case in order to make the solution free of singularities. Some of these works include the regularization of the Reisnerr-Nordstr\"om and Kerr-Newman black holes \cite{Simpson:2021vxo}, thin disc accretion analysis for the Simpson-Visser solution \cite{Bambhaniya:2021ugr}, the possibility of the existence of traversable wormholes in semiclassical gravity \cite{Terno:2022qot}, solutions in modified gravities \cite{Junior:2022fgu}, gravitational lensing \cite{Islam:2021ful, Nascimento:2020ime, Tsukamoto:2020bjm},  the formation of ``shadows'' around the central object  \cite{Guerrero:2021ues}, the use of the Gauss-Bonet theorem to determine light deflection and the analysis of the effects of dark matter \cite{Ovgun:2020yuv}, the need of phantom fields \cite{Chataignier:2022yic, Bronnikov:2022bud}, among other works \cite{Stuchlik:2021tcn, Churilova:2019cyt, Yang:2022ryf, Vagnozzi:2022moj, Lobo:2020ffi}. One significant question that arises is the nature of the source that could generate such solutions, as they require exotic sources. In the literature, as discussed in \cite{Rodrigues:2023vtm, Bronnikov:2021uta}, it is quite common to use nonlinear electrodynamics (NED) to describe regular black hole solutions, including those with multiple horizons. Another common approach involves a phantom field to obtain traversable wormholes, as they require exotic matter that violates the null energy condition (NED). However, these individual sources cannot be used to describe "black bounce" solutions. In the case of NED, the condition $T^0{}_0=T^1{}_1$ must hold, which is clearly violated in the case of the present paper and in other black bounce scenarios. Similarly, for a regular scalar field or a phantom field, the condition $T^0{}_0=T^2{}_2=T^3{}_3$ should be satisfied, which is also not valid for black bounce solutions. One way to circumvent this problem is to assume a coupling between  two types of solutions, where the presence of a phantom field is expected due to wormholes being part of the Simpson-Visser solution, while NED complements the solution to set the stress-energy tensor to the Simpson-Visser spacetime, representing regular black holes \cite{Bronnikov:2021uta}. Extensive scientific literature can also be found by using these sources for black strings, as in \cite{Bakhtiarizadeh:2023mhk}.

Recently, several works have been published where this regularization is applied to solutions with cylindrical symmetry \cite{Furtado:2022tnb, Lima:2022pvc, Bronnikov:2023aya, Lima:2023arg}, known as black strings. In these cases, the regularization $r^2\rightarrow r^2+a^2$ was applied to metrics defined for static black strings \cite{Lemos:1994xp, Lemos:1995cm} and the  results were extremely similar to the cases of static black bounce solutions with spherical symmetry concerning the interpolation between black hole and traversable wormhole solutions and the violation of the null energy condition for all types of solutions inside and outside the event horizon, as well as being regular for every value of $a\neq 0$. However, the solution for rotating black strings, as described in \cite{Lemos:1995cm}, which are analogous to the case of the regularized Kerr or Kerr-Newman solution, have not yet been explored yet.

Hence, the aim of the present paper is to apply the Simpson-Visser procedure to a rotating black string, following similar previous analysis \cite{Franzin:2021vnj}, and study the new features that might arise for this type of solution. To do so, we study the structure of the new solution, where new feature are found as well as its regularity by analysing the different curvature magnitudes. Finally, through the Einstein's field equations, the corresponding energy conditions are obtained for an effective energy-momentum tensor.   

The paper is organised as follows: in section \ref{Sec-2} the new solution for a regular black string is obtained. Section \ref{EHSG} is devoted to the analysis of the structure of this new metric. In section \ref{CI}, the corresponding curvature invariants are obtained and show the regularity of the solution. Section \ref{EC} shows the energy-conditions. Finally, section \ref{conclusions} gathers the main results of the paper.


\section{Black-bounce for rotating black strings}\label{Sec-2}


Let us start by introducing the spacetime metric that describes a rotating black string in a form analogous to the spherically symmetric case expressed in Boyer-Lindquist coordinates \cite{Lemos:1995cm}:
\begin{eqnarray}\label{1}
     ds^2=-\Delta_{NR}(r)\left(\gamma dt-\frac{\omega}{\alpha^2}d\varphi\right)^2+r^2(\gamma d\varphi-\omega dt)^2+\frac{dr^2}{\Delta_{NR}(r)}+\alpha^2r^2dz^2\ , 
\end{eqnarray}
where
\begin{eqnarray}\label{2}
    \Delta_{NR}(r)=\alpha^2r^2-\frac{b}{\alpha r};\, b=4M\left(1-\frac{3J^2\alpha^2}{2}\right);\, \gamma=\sqrt{\frac{1-J^2\alpha^2/2}{1-3J^2\alpha^2/2}};\, \omega=\frac{J\alpha^2}{\sqrt{1-3J^2\alpha^2/2}}\ . 
\end{eqnarray}
Here $\alpha^2=-\Lambda/3$ is an effective cosmological constant, $M$ is the mass density and $J$ is a constant associated with the angular momentum density. For faraway observers, this solution extends uniformly along the z-axis, providing a $\mathcal{R}\times\mathcal{S}^1$ topology. Note also that this spacetime is not asymptotically flat but anti-de Sitter. As shown in Ref.~\cite{Lemos:1995cm}, this theoretical spacetime can be generated by using nonlinear electrodynamics (NED) and a scalar field, which will also include a non-null electric charge, not considered here in order to simplify the mathematical expressions (its extension can be followed trivially). \\

Nevertheless, the spacetime metric (\ref{1}) is geodesically incomplete, since it contains a singularity located at $r=0$ as can be easily inferred by analysing the Kretschmann scalar: 
\begin{equation}\label{3}
    R^{\mu\nu\lambda\rho}R_{\mu\nu\lambda\rho} = 24\alpha^4\left(1+\frac{b^2}{2\alpha^6r^6}\right).
\end{equation}
As in the Kerr metric, the singularity has a ring structure. Our goal here is to regularize the above metric (\ref{1}) by introducing a bounce on the radial coordinate, such that the Kretschmann scalar becomes regular at every spacetime point. This follows from the Simpson-Visser spherically symmetric case \cite{Simpson:2018tsi}, also implemented successfully for the Kerr-Newman metric \cite{Franzin:2021vnj}. To do so, one assumes the radial coordinate in the metric as $r^2\rightarrow r^2+a^2$. Then, the metric (\ref{1}) turns out:
\begin{eqnarray}\label{4}
     ds^2=-\Delta(r)\left(\gamma dt-\frac{\omega}{\alpha^2}d\varphi\right)^2+(r^2+a^2)(\gamma d\varphi-\omega dt)^2+\frac{dr^2}{\Delta(r)}+
    \alpha^2(r^2+a^2)dz^2, 
\end{eqnarray}
where
\begin{equation}\label{5}
    \Delta(r)=\alpha^2(r^2+a^2)-\frac{b}{\alpha\sqrt{r^2+a^2}}.
\end{equation}
One can easily note that this metric is regular at $r=0$ and also the Kretschmann scalar is. The radial coordinate is actually defined in the range $r\in (-\infty, +\infty)$ and nothing occurs at $r=0$ but the cylinder has a minimum size that corresponds to a throat connecting two universes (for more details, see below). It is straightforward to verify that this solution reduces to the one defined in (\ref{1}) as $a\rightarrow 0$. This result is also consistent with the regularization of a static and neutral black string carried out in \cite{Lima:2022pvc} by setting $J=0$, which results in $\omega=0$, $\gamma=1$, and $b=4M$. In the following, we study the properties of this metric by analyzing event horizons, surface gravity, curvature quantities and finally by evaluating the energy-momentum tensor through the Einstein equations in order to assess the energy conditions of the system.

\section{Event Horizons, Surface Gravity and Ergoregion} \label{EHSG}

To study the properties of the black string described by the metric (\ref{4}), let us discuss the possible existence of horizons and ergospheres. which follow the approach used in black-bounce works by evaluating the metric defined in (\ref{5}). First, we will assess the existence of event horizons and examine the behaviour of this type of solution as a function of the Simpson-Visser parameter $a$. The horizons are given by the roots of $\Delta(r)$, which define null hypersurfaces:
\begin{equation}\label{7}
\Delta(r)=0\rightarrow r_H=\pm \sqrt{\frac{b^{2/3}}{\alpha^2}-a^2}=\pm\sqrt{\frac{1}{\alpha^2}\left[16M^2\left(1-\frac{3J^2\alpha^2}{2}\right)\right]^{1/3}-a^2}\ ,
\end{equation}
where we have used the expressions in (\ref{2}). Note that $r_H=b^{1/3}/\alpha$ corresponds to the position of the event horizon for the standard solution ($a=0$). Hence, depending on the relative value of the parameter $a$, one might have the following possibilities:
\begin{itemize}
\item For $a^2>\frac{b^{2/3}}{\alpha^2}$. For this case, the equation $\Delta(r)=0$ has no real roots, such that there is no event horizons and the metric (\ref{4}) describes a rotating cylindrically symmetric traversable wormhole.
\item For $a^2<\frac{b^{2/3}}{\alpha^2}$. The equation $\Delta(r)=0$ has two real roots that corresponds to two cylindrically symmetric horizons, each one located in one universe. The metric (\ref{4}) describes a regular black hole.
\item For $a^2=\frac{b^{2/3}}{\alpha^2}$. The unique solution for the equation $\Delta(r)=0$ is $r=0$, such that there is just one event horizon that coincides with the throat of the wormhole. This is analogous to the spherically symmetric case discussed in Ref.~\cite{Simpson:2018tsi} and the metric (\ref{4}) corresponds to a non-traversable cylindrically symmetric rotating wormhole.
\end{itemize}
We observe that these results are consistent with black-bounce type solutions, where the horizon position, if any, is corrected by a factor of the form $\sqrt{1-a^2/r_{HNR}^2}$, with $r_{HNR}$ being the position of the horizon in the usual black string solution. Hence, the result in (\ref{7}) is consistent with previous results \cite{Lemos:1994xp} when setting $a=0$ and identical to that in \cite{Lima:2022pvc} for $J=0$. Moreover, 


The surface gravity can be also easily obtained as a function of $r_H$ in those cases that exists:
\begin{equation}
\kappa_{r_H}=\frac{1}{2}\frac{d}{dr}\Delta(r_H)=\kappa_{NR}\sqrt{\frac{r_H^2}{r_H^2+a^2}}\ ,
\end{equation}
where $\kappa_{NR}=\frac{3\alpha b^{1/3}}{2}$ is the surface gravity of the standard solution. This result is again consistent with the standard case ($a=0$) and with the regular static solution ($J=0$).\\

Moreover, the metric (\ref{4}) might also provide an ergoregion, limited by the roots of $g_{tt}=0$, which are given by:
\begin{equation}
r_{E}=\pm \sqrt{\left[\frac{b\gamma^2}{\alpha (w^2-\alpha^2\gamma^2)}\right]^{2/3}-a^2}=\pm\sqrt{\frac{1}{\alpha^2}\left[16M^2\left(1-\frac{J^2\alpha^2}{2}\right)\right]^{1/3}-a^2}\ .
\label{7b}
\end{equation}
As above, this open a new set of possibilities, absent in the singular black string solution (\ref{1}). In addition, note that for this type of metrics, the ergoregion will show up as an infinite cylinder, as the topology of the event horizon. Moreover, unlike the Kerr spacetime, both hypersurfaces do not coincide at any point. Let us define $B=\left[16M^2\left(1-\frac{J^2\alpha^2}{2}\right)\right]^{1/3}$, then the following possibilities might arise: 
\begin{itemize}
\item For $a^2\geq B$, there is no ergoregion and neither an event horizon. 
\item For $a^2<B$, there is an ergoregion. The possible existence of an event horizon depends on the relative value of $a$, as explained above. 
\end{itemize}
Now that we have assessed some basic properties of the metric, we need to study the curvature quantities to analyze the regularity of this solution. For this purpose, we will determine the curvature invariants such as the Ricci scalar and the Kretschmann scalar, as well as the curvature tensors such as the Riemann tensor and the Ricci tensor, and evaluate them at $r=0$ to verify their finiteness.

\section{Curvature Invariants and regularity}
\label{CI}

To evaluate whether this solution is free of singularities, the analysis of the behaviour of curvature magnitudes evaluated at $r=0$ is required, since the original metric shows a singularity, similar to the one in the Kerr metric. For the case of static solutions, the analyze of the Kretschmann scalar is enough, as pointed out in Refs.~\cite{Lima:2022pvc, Lima:2023arg, Franzin:2021vnj}. Wether this scalar is finite when evaluated at $r=0$, the solution turns out regular \cite{Simpson:2023apa}. Nevertheless, since the solution studied here is not static but rotating, the Kretschmann scalar by itself is not enough for the regularity analysis, which means that a complete analysis including other curvature invariants is necessary \cite{Franzin:2021vnj}.

Then, the Ricci scalar, the Ricci contraction and the Kretschmann scalar can be easily obtained for the spacetime metric defined in (\ref{5}), leading to:
\begin{eqnarray}
    &&R=-12\alpha^2+\frac{3\alpha^2a^2[b/\alpha^3+2(r^2+a^2)^{3/2}]}{(r^2+a^2)^{5/2}}; \label{9}\\
     &&R^{\mu\nu}R_{\mu\nu}=36\alpha^4-\frac{18a^2\alpha^4[b/\alpha^3+2(r^2+a^2)^{3/2}]}{(r^2+a^2)^{5/2}}+ 
    \frac{3a^4\alpha^4[3b^2/\alpha^6+4(r^2+a^2)^{3/2}b/\alpha^3+8(r^2+a^2)^3]}{2(r^2+a^2)^5}; \label{10}\\
     &&R^{\mu\nu\lambda\rho}R_{\mu\nu\lambda\rho} = 24\alpha^4\left(1+\frac{b^2}{2\alpha^6(r^2+a^2)^3}\right)+
    \frac{3a^2b^2[11a^2-12(r^2+a^2)]}{\alpha^2(r^2+a^2)^5}+
    \frac{12\alpha^4a^2[a^2\sqrt{r^2+a^2}-2(r^2+a^2)^{3/2}-b/\alpha^3]}{(r^2+a^2)^{5/2}}. \label{11}
\end{eqnarray}
One can note that these results are consistent with the standard black string solution by setting $a=0$, as shown in \cite{Hendi:2013mka}, and  are consistent with the case of a regular neutral black string when $J=0$, as shown in \cite{Lima:2022pvc}. By evaluating the above expressions (\ref{11}) at $r=0$, one finds:
\begin{eqnarray}
    &&R=-6\alpha^2+\frac{3b}{\alpha a^3};\\
    &&R^{\mu\nu}R_{\mu\nu}=12\alpha^4+\frac{9b^2}{2\alpha^2a^6}-\frac{12\alpha b}{a^3};\\
    &&R^{\mu\nu\lambda\rho}R_{\mu\nu\lambda\rho} = 24\alpha^4\left(\frac{1}{2}+\frac{3b^2}{8\alpha^6a^6}-\frac{b}{2\alpha^3a^3}\right).
\end{eqnarray}
We can easily infer that this solution is free of singularities as far as $a\neq 0"$, which is the fundamental parameter for regularizing the solution, as these results show. Let us now complete this analysis on the regularity of the solution (\ref{5}) by evaluating the curvature tensors at $r=0$ and verify that all the components of these tensors are finite there. Since our metric is not orthogonal, the tensors have a more complex form than the static case. For instance, the Ricci tensor is not diagonal as the component $R_{02}$ is non-zero (as neither $R_{20}$). One way to simplify the analysis is to work in the local frame by using the so-called "orthonormal tetrads" (for more details on the local frame, one can study Cartan structures in \cite{mcmahon:relativity2006}). To simplify the calculations, let's perform a change of variable in $r$ by defining the new radial variable $x$ such that $x^2=r^2+a^2$, and $dr^2=(1-a^2/x^2)^{-1}dx^2$. In terms of this new variable, the orthonormal tetrads can be written as:
\begin{eqnarray}
    (e_{\hat{0}})^{\mu}&=&\frac{\alpha^2}{\sqrt{|\Delta|}(\alpha^2\gamma^2-\omega^2)}(\gamma,\, 0,\, \omega,\, 0);\\
    (e_{\hat{1}})^{\mu}&=&\sqrt{|\Delta|\left(1-\frac{a^2}{x^2}\right)}(0,\, 1,\, 0,\, 0);\\
    (e_{\hat{2}})^{\mu}&=&\frac{1}{x(\alpha^2\gamma^2-\omega^2)}(\omega,\, 0,\, \alpha^2\gamma,\, 0);\\
    (e_{\hat{3}})^{\mu}&=&\frac{1}{\alpha x}(0,\, 0,\, 0,\, 1).
\end{eqnarray}
This result is consistent with what one would expect for the non-rotating case as the terms where the index $\mu$ is different from the index with the "hat" vanish when we set $\omega=0$. Therefore, the tetrads are diagonal, as occurs for the static solution. In addition, for $\gamma=1$ the terms of the tetrads for the static case are recovered, according to the metric defined in \cite{Lima:2022pvc}.

Hence, from these results, we can determine the components of the curvature tensors. In particular, we can obtain the Ricci tensor as follows:
\begin{eqnarray}
     &&R_{\hat{0}\hat{0}} = \text{sign}[\Delta]\left[3\alpha^2+\frac{a^2\alpha^2(b/\alpha^3-4x^3)}{2x^5}\right]; \\
    &&R_{\hat{1}\hat{1}} = \text{sign}[\Delta]\left[\frac{3a^2\alpha^2(b/\alpha^3)}{2x^5}-3\alpha^2\right]; \\
    &&R_{\hat{2}\hat{2}} = R_{\hat{3}\hat{3}} = \frac{a^2\alpha^2}{x^2}\left(\frac{b}{\alpha^3x^3}+2\right)-3\alpha^2,
\end{eqnarray}
where $\text{sign}[\Delta]=\pm 1$, with the positive sign used when $\Delta>0$, i.e., outside the event horizon, and the negative sign is used when $\Delta<0$, i.e., inside the horizon. Note that all these components are identical to those of the regular neutral black string when written as $R^{\hat{\mu}\hat{\nu}}$, explicitly showing their consistency with this solution when $J=0$. It can also be observed that all of them, in the form of a $(1,1)$-tensor result in $-3\alpha^2$ for $a=0$, as expected since the vacuum solution is recovered, where $R^{\hat{\mu}\hat{\nu}}=\Lambda=-3\alpha^2$.

To evaluate these components at $r=0$, one has to take $x=a$, according to the definition for $x$. Then, the following expressions for the Ricci components are obtained:
\begin{eqnarray}
     &&R_{\hat{0}\hat{0}} = \text{sign}[\Delta]\left[\alpha^2+\frac{b}{2\alpha a^3}\right]; \\
    &&R_{\hat{1}\hat{1}} = \text{sign}[\Delta]\left[\frac{3b}{2\alpha a^3}-3\alpha^2\right]; \\
    &&R_{\hat{2}\hat{2}} = R_{\hat{3}\hat{3}} = \frac{b}{\alpha a^3}-\alpha^2.
\end{eqnarray}
Hence, as we have shown above, all these components are finite as far as $a\neq 0$. The same applies to the Riemann tensor, which also has no divergences at $r=0$ under this condition. Therefore, we can ensure that the spacetime metric (\ref{5}) is indeed regular, similar to what happens in the case of Kerr-Newman black hole, except for the fact that in this solution, $R_{\hat{2}\hat{2}} = R_{\hat{3}\hat{3}}$, what does not occur in the regularization of Kerr-Newman black hole. In other words, while the Kerr-Newman solution has all distinct pressures, representing a completely anisotropic fluid, our solution is anisotropic with $p_{\varphi}=p_{z}$, analogous to what occurs in other black string bounce solutions.

In the next section, the corresponding energy conditions associated with the stress-energy tensor are analyzed through Einstein equations field equations.

\section{Stress-Energy Tensor and Energy Conditions}\label{EC}

In order to obtain the corresponding effective energy-momentum tensor, we follow the Einstein equations, given by $G_{\hat{\mu}\hat{\nu}}+\eta_{\hat{\mu}\hat{\nu}}\Lambda=8\pi T_{\hat{\mu}\hat{\nu}}$. Then, the components of the Einstein tensor from the Ricci tensor and the Ricci scalar are:
\begin{eqnarray}
     &&G_{\hat{0}\hat{0}} = \text{sign}[\Delta]\left[\frac{a^2\alpha^2(2b/\alpha^3+x^3)}{x^5}-3\alpha^2\right]; \\
    &&G_{\hat{1}\hat{1}} = \text{sign}[\Delta]3\alpha^2\left(1-\frac{a^2}{x^2}\right); \\
    &&G_{\hat{2}\hat{2}} = G_{\hat{3}\hat{3}} = 3\alpha^2-\frac{a^2\alpha^2}{x^2}\left(\frac{b}{2\alpha^3x^3}+1\right).
\end{eqnarray}
This result is once again identical to what was for the regular neutral black string ($J=0$) when considering the Einstein tensor and it reproduces the usual case for $a=0$, where these components should all be equal to $-\Lambda=3\alpha^2$.

On the other hand, the components of the stress-energy tensor, outside the event horizon (when $x>b^{1/3}/\alpha$), are given by:
\begin{equation}\label{29}
    T_{\hat{0}\hat{0}}=\rho;\, T_{\hat{1}\hat{1}}=p_{\|};\, T_{\hat{2}\hat{2}}=T_{\hat{3}\hat{3}}=p_{\bot}.
\end{equation}
Hence, outside the event horizon, for example, we will have:
\begin{eqnarray}
    &&\rho = \frac{\alpha^2a^2(2b/\alpha^3+x^3)}{8\pi x^5}; \label{30}\\
    &&p_{\|} = -\frac{3\alpha^2a^2}{8\pi x^2}; \label{31}\\
    &&p_{\bot} = -\frac{a^2\alpha^2}{8\pi x^2}\left(\frac{b}{2\alpha^3x^3}+1\right). \label{32}
\end{eqnarray}
To determine the components of the energy-momentum tensor inside the horizon, we simply set $\text{sign}[\Delta]=-1$ and interchange the spacelike and timelike features. In other words, we will have $T_{\hat{0}\hat{0}}=p_{|}$ and $T_{\hat{1}\hat{1}}=\rho$, as mentioned in \cite{Simpson:2018tsi}.

Now let's examine the energy conditions. According to \cite{Bronnikov:2021uta}, the null energy condition (NEC) is necessarily violated in black bounce solutions. Consequently, the weak energy condition (WEC) is also violated, as the WEC is equivalent to the NEC with the additional condition $\rho\geq 0$. However, the strong energy condition (SEC) and the dominant energy condition (DEC) are not always violated. Let's specifically evaluate the NEC and the SEC:
\begin{eqnarray}
    &&(\text{NEC})\rightarrow \rho + p_{\|} = \text{sign}[\Delta]\left[\frac{2a^2\alpha^2(b/\alpha^3-x^3)}{8\pi x^5}\right]; \label{33}\\ 
     &&(\text{SEC})\rightarrow \rho + p_{\|} + 2 p_{\bot} = 
    \text{sign}[\Delta]\left[\frac{2a^2\alpha^2(b/\alpha^3-x^3)}{8\pi x^5}\right]-
    \frac{a^2\alpha^2(b/\alpha^3+2x^3)}{8\pi x^5}. \label{34}
\end{eqnarray}
By analyzing the NEC, we see that it is indeed violated both inside and outside the horizon. Note that in the coordinate $x$, the position of the horizon is given by $b^{1/3}/\alpha$. Hence, outside the horizon, one should have $b/\alpha^3-x^3<0$ and $\text{sign}[\Delta]=1$, which makes the expression (\ref{30}) negative for every $x$. Inside the horizon,  $b/\alpha^3-x^3>0$ and $\text{sign}[\Delta]=-1$ hold, such that it results again in a negative value. As for the SEC, each case has to be evaluated separately:
\begin{eqnarray}
    &&\rho + p_{\|} + 2 p_{\bot} = \left[\frac{a^2\alpha^2(b/\alpha^3-4x^3)}{8\pi x^5}\right],\, \text{se}\, \Delta>0; \label{35}\\
    &&\rho + p_{\|} + 2 p_{\bot} = -\frac{3ba^2}{8\pi \alpha x^5},\, \text{se}\, \Delta<0. \label{36}
\end{eqnarray}
Since $b/\alpha^3-4x^3<0$ for $x>b^{1/3}/\alpha$, the SEC must also be violated for every $x$ inside and outside the event horizon.

\subsection{Sources for rotating black strings bounces}
Finally, we will explore the possible sources for the rotating black-bounce solution (\ref{4}). As shown in previous literature \cite{Rodrigues:2023vtm, Bronnikov:2021uta, Bronnikov:2023aya}, static black strings regularized by following the Simpson-Visser procedure can be generated through a combination of a phantom scalar field and a source associated with nonlinear electrodynamics (NED). Both sources are necessary, since it's not possible to consider just one of them. This stems from the fact that by considering only the scalar field, one obtains $T^0{}_0=T^2{}_2$, which is not true for static black-bounce solutions. On the other hand, by assuming only NED leads to $T^0{}_0=T^1{}_1$, which is also not valid for these solutions. However, when both sources are considered together, a final solution is obtained with energy-mass density independent of the radial pressure and both also independent of lateral pressures.

Regarding the source associated with electrodynamics, one can choose an electric source with the presence of only electric charge or a magnetic source with the presence of a magnetic charge. Both types of systems are capable of generating regular solutions for static black strings. However, the electric sourced solution is unable to recover the behavior of Maxwell's electrodynamics in the weak-field limit ($L_F\rightarrow 1$). In the case of dyonic solutions, where both charges exist simultaneously, they cannot generate a regular black hole solution. Further details can be found in \cite{Bronnikov:2022ofk}.

What we are going to attempt here is to consider the same sources for the rotating case as well by assuming the following action:
\begin{equation}\label{37}
    S=\int d^4x \sqrt{-g}[R-2\Lambda - 16\pi (\epsilon g^{\mu\nu}\partial_{\mu}\phi\partial_{\nu}\phi+V(\phi))-16\pi L(F)].
\end{equation}
Here $\phi(r)$ represents the scalar field that is coupled to the system, which can be a canonical scalar field when setting $\epsilon=1$ (positive kinetic energy) or a phantom field for $\epsilon=-1$ (negative kinetic energy), while $V(\phi)$ is the potential energy associated to the scalar field. The Lagrangian $L(F)$ describes a nonlinear electrodynamics, where $F=F^{\mu\nu}F_{\mu\nu}/4$ represents the usual Maxwell Lagrangian. By applying the variational principle to the action in (\ref{37}), the following equations of motion are obtained:  
\begin{eqnarray}
    &&R_{\mu\nu}-\frac{1}{2}g_{\mu\nu}R+g_{\mu\nu}\Lambda=8\pi([T_{\phi}]_{\mu\nu} +[T_{NED}]_{\mu\nu}), \label{38}\\
    &&2\epsilon \nabla_{\mu}\nabla^{\mu}\phi=\frac{dV}{d\phi}, \label{39}\\
    &&\nabla_{\mu}[L_F F^{\mu\nu}]=\frac{1}{\sqrt{-g}}\partial_{\mu}[\sqrt{-g}L_F F^{\mu\nu}]=0, \label{40}\ ,
\end{eqnarray}
where $L_F=\partial L/\partial F$ and the components of the scalar field and NED stress-energy tensor are respectively:
\begin{eqnarray}
    &&[T_{\phi}]_{\mu\nu}=2\epsilon\partial_{\mu}\phi\partial_{\nu}\phi-g_{\mu\nu}(\epsilon\partial^{\alpha}\phi\partial_{\alpha}\phi+V(\phi)), \label{41}\\
    &&{}[T_{NED}]_{\mu\nu}=L_F F_{\mu}{}^{\alpha}F_{\nu\alpha}-g_{\mu\nu}L(F). \label{42}
\end{eqnarray}
Following Ref.~ \cite{Bronnikov:2023aya}, we will assume the presence of a magnetic charge $Q$ as the source of the electrodynamics part, such that $F_{23}=-F_{32}=Q$. Consequently, the Maxwell invariant leads to:
\begin{equation}\label{43}
    F=\frac{Q^2\alpha^2\gamma^2}{2x^4(\alpha^2\gamma^2-\omega^2)^2}+\frac{Q^2\alpha^3\omega^2}{2x(b-\alpha^3x^3)(\alpha^2\gamma^2-\omega^2)^2}.
\end{equation}
One can note that in absence of rotation ($J=0$), which implies $\omega=0$ and $\gamma=1$, one leads to $F=Q^2/2x^4\alpha^2$, as was found in Ref.~\cite{Bronnikov:2023aya}. We can interpret $Q$ as a magnetic charge density per length, so the term $F$ will have the units of magnetic charge squared per length to the fourth power. 


Hence, the non-zero components of the energy-momentum tensor in the local frame are:
\begin{eqnarray}
    &&T_{\hat{0}\hat{2}}=T_{\hat{2}\hat{0}}=\frac{Q^2\alpha^2\omega\gamma L_F(r(x))}{x^3\sqrt{\alpha^2x^2-\frac{b}{\alpha x}}(\alpha^2\gamma^2-\omega^2)^2};\label{44}\\
    &&T_{\hat{0}\hat{0}}=\frac{Q^2\alpha^3\omega^2L_F(r(x))}{x(\alpha^3x^3-b)(\alpha^2\gamma^2-\omega^2)^2}+\frac{(a^2-x^2)(b-\alpha^3x^3)\epsilon\phi'(r(x))^2}{\alpha x^3}+V(r(x))+L(r(x));\label{45}\\
    &&T_{\hat{1}\hat{1}}=\frac{(a^2-x^2)(b-\alpha^3x^3)\epsilon\phi'(r(x))^2}{\alpha x^3}- V(r(x))-L(r(x));\label{46}\\
    &&T_{\hat{2}\hat{2}}=\frac{Q^2\alpha^2\gamma^2L_F(r(x))}{x^4(\alpha^2\gamma^2-\omega^2)^2}+\frac{(a^2-x^2)(b-\alpha^3x^3)\epsilon\phi'(r(x))^2}{\alpha x^3}-L(r(x))-V(r(x));\label{47}\\
    &&T_{\hat{3}\hat{3}}=\frac{Q^2\alpha^2\gamma^2L_F(r(x))}{x^4(\alpha^2\gamma^2-\omega^2)^2}+\frac{(a^2-x^2)(b-\alpha^3x^3)\epsilon\phi'(r(x))^2}{\alpha x^3}-L(r(x))-V(r(x))-\frac{Q^2\alpha^3\omega^2}{x(\alpha^3x^3-b)(\alpha^2\gamma^2-\omega^2)^2}\ .\label{48}
\end{eqnarray}
Here the primes denote derivatives with respect to $r$. However, some inconsistencies arise when comparing these expressions with the ones obtained above in (\ref{30}), (\ref{31}), and (\ref{32}), since for these sources, we have $T_{\hat{2}\hat{2}}\neq T_{\hat{3}\hat{3}}$, although both lateral pressures must remain the same. Furthermore, the component $T_{\hat{0}\hat{2}}$ is not zero, which is not reasonable, as in the local frame, the energy-momentum tensor should be diagonal, as obtained above. However, when considering the static case ($J=0$), the result becomes entirely consistent with the solution, as expression (\ref{44}) reduces to zero and $T_{\hat{2}\hat{2}}=T_{\hat{3}\hat{3}}$. We can conclude, therefore, that the combination of a scalar field with a magnetic charge associated with NED is not enough to generate a rotating black string solution, even though it can reproduce static solutions.

One way to overcome this issue might be to consider a third source in such a way that makes the equations consistent. To do so, an anisotropic fluid is necessary, whose components of the energy-momentum tensor associated to the $\hat{\varphi}$ and $\hat{z}$ coordinates must be different in order to accomplish $T_{\hat{2}\hat{2}}=T_{\hat{3}\hat{3}}$ when added to the total energy-momentum tensor in (\ref{47}) and (\ref{48}). Furthermore, the component $T_{\hat{0}\hat{2}}$ of this fluid must cancel out the term (\ref{44}), resulting in a complete system compatible with the solution (\ref{4}). This method would be analogous to what was done in \cite{Franzin:2021vnj}, in the section where the authors discuss the interpretation of the charge-dependent part of the energy-momentum tensor (here considering the regularization of Kerr-Newman, a charged and rotating solution), and among the interpretations, they consider an anisotropic fluid.

Regarding the scalar field, we have the equation:
\begin{equation}\label{49}
    (a^2-x^2)(b-\alpha^3x^3)\phi''(r(x))\phi'(r(x))-x[b+\alpha^3x(3a^2-4x^2)]\phi'(r(x))^2=\frac{\alpha x^3V'(r(x))}{2\epsilon}.
\end{equation}
We can solve it to find $\phi(r(x))$, which in principle should be possible to be integrated together with the Einstein field equations.


\section{Conclusions}\label{conclusions}

Along this paper we have considered the solution found in Ref.~\cite{Lemos:1995cm} for a rotating black string and have found a new regular solution motivated by the Simpson-Visser black bounce. Then, we have studied the structure of the spacetime metric by analyzing the existence of event horizons and ergoregions, revealing the same pattern as in the case of standard black bounce solutions, where there is an interpolation between a regular black hole and a traversable wormhole. The surface gravity of this solution was also obtained, showing consistency with some previous results found for the singular black string case as well as for the static metric \cite{Lima:2022pvc}.

To verify the regularity of the solution, an analysis of the curvature magnitudes has been conducted, particularly focusing on the curvature invariants such as the Ricci and Kretschmann scalars and the Ricci tensor. All non-zero components of the Ricci tensor were also determined, using the non-holonomic basis and the "tetrads" method to diagonalize the Ricci and Einstein tensors. By evaluating these quantities at $r=0$, all of them become finite and consequently free of singularities, as long as $a\neq 0$, confirming the solution's regularity. 

Then, by the Einstein field equations, we have obtained the components of the effective energy-momentum tensor for this solution. By analyzing its components, we have obtained the energy conditions, showing that the null and strong energy conditions are violated, which follows the same pattern as in other  black bounce solutions where the null energy condition is necessarily violated.

Regarding the energy-momentum tensor, we have explored the possible potential sources for this type of solution based on previous analysis regarding sources for static black bounces. These sources involve a scalar field and nonlinear electrodynamics with a magnetic source. By following a similar approach as in Ref.~\cite{Bronnikov:2023aya} for the static case of regular black strings, we found that the rotating case becomes a challenge to find a consistent energy-momentum tensor that satisfies the Einstein field equations. This is due to the presence of three different pressures and a component off the diagonal in the tensor's matrix representation. To address this issue, we can introduce a third source that is provided by an anisotropic fluid, which makes the solution consistent with the rotating scenario and forms a basis for future research on black bounce sources with rotation. However, this analysis requires a bit more attention and therefore will be carried out in future works.

For a future work, we lead the analysis of the orbits that particles might follow within this spacetime, which might be fundamental to understand the possibilities of having such type of solutions as viable solutions for real compact objects, at least effectively.

\section*{Acknowledgements}

The authors would like to thank Conselho Nacional de Desenvolvimento Científico e Tecnológico (CNPq), Fundação Cearense de Apoio ao Desenvolvimento Científico e Tecnológico
(FUNCAP) and Coordenação de Aperfeiçoamento de Pessoal de Nível Superior - Brasil (CAPES) for finantial support. DS-CG is supported by Spanish grant Ref.~PID2020-117301GA-I00 funded by MCIN/AEI/10.13039/501100011033 (``ERDF A way of making Europe" and ``PGC Generaci\'on de Conocimiento").

\end{document}